# Controlled integration of MoS$_2$ flakes on nanopores by means of electrophoretic deposition

*Dario Mosconi[1], Andrea Jacassi[2], Giorgia Giovannini[2], Paolo Ponzellini[2], Nicolò Maccaferri[2], Paolo Vavassori[3], Michele Dipalo[2], Francesco De Angelis[2], Stefano Agnoli[1], and Denis Garoli[2\**]*

[1] Dipartimento di Chimica, Università degli Studi di Padova, Via Marzolo 1, 35131, Padova, Italy.
[2] Istituto Italiano di Tecnologia – Via Morego, 30, I-16163 Genova, Italy
[3] CIC nanoGUNE - Tolosa Hiribidea, 76, E-20018 Donostia – San Sebastian, Spain
\* Corresponding author's email: denis.garoli@iit.it.

## Abstract

We propose an easy and robust strategy for the versatile integration of 2D material flakes on plasmonic nanoholes by means of controlled deposition of MoS$_2$ via electrophoretic process. The method can be applied both to simple metallic flat nanostructures and to complex 3D metallic structures both comprising nanoholes. The deposition method allows the decoration of large ordered arrays of plasmonic structures with single or few layers of MoS$_2$. We show that the plasmonic field generated by the nanohole can interact significantly with the 2D layer, thus representing an ideal system for hybrid 2D-Material/Plasmonic investigation. The controlled and ordered integration of 2D materials on plasmonic nanostructures opens a pathway towards, for instance, enhanced light emission; strong coupling from plasmonic hybrid structures; hot electron generation; and sensors based on 2D materials.

## Introduction

In the last decade, intensive research efforts have been devoted to the investigation of two-dimensional (2D) materials, such as graphene and various transition metal chalcogenides (TMDCs).[1–5] Molybdenum disulfide (MoS$_2$) is a typical representative of TMDCs materials family, which consists of S-Mo-S layers bonded by van der Waals interaction. MoS$_2$ offers a powerful platform for applications in nanophotonics and optoelectronics due to its remarkable optical properties. In fact, MoS$_2$ behaves like a semiconductor with direct band gap electronic structure, notable flexibility, and tunable optical emission.[6,7] Owing to its atomically thin layer, monolayer MoS$_2$ can be easily integrated with other low-dimensional materials such as quantum dots,[8] nanowires,[9,10] and other 2D materials[11,12] to form hybrid nanostructures with intriguing electronic and optical properties.

Regarding the latter, an extensive interest from researchers raised in the last years to explore potential physical phenomena, especially on light−matter interaction, such as multiexciton photoluminescence (PL), interlayer exciton coupling, strong coupling and valley polarization effect.[13–17] However, enhancing and modulating the quantum optical properties of $MoS_2$ layer(s) are still a great challenge for its practical applications. Great efforts have been devoted to realize the active control of $MoS_2$ by utilizing photonic cavity modes and metallic surface plasmons.[18–23] However, most designed structures must be fabricated precisely with the structure geometries and the coupling distance, which limits the development of active-controlled optoelectronic devices in the future. Moreover, one of the main limitation of $MoS_2$ application in photonics and plasmonics is related to the challenging alignment between nanostructures and the 2D material. New approached for easy and controlled integration of 2D materials with metallic or semiconductor nanostructures can now represent a significant help in the cited field of research. We have recently proposed and demonstrated the possibility to control the deposition of single layer $MoS_2$ flakes on metallic nanostructures by means of chemical conjugation.[24] Here we propose an alternative approach that can be in principle applied to any substrates comprising nanoapertures to be decorated with $MoS_2$ flakes as single or few layers. As we will show, the procedure is based on electrophoretic deposition where the substrate do not act as electrode. The method allows high yield controlled deposition in both flat nanostructures and 3D elements.

We think that the structures that can be easily prepared with this method can find several interesting applications in all the present fields of research in which 2D materials are the core. For example, the controlled/ordered integration of 2D materials on plasmonic nanostructures can pave the way to new investigations on enhanced light emission from TMDCs,[13,21,25–30] strong coupling from plasmonic hybrid structures,[17] hot electron generation,[31,32] and sensors in general based on 2D materials.[33–38]

**Results and Discussion**

The deposition process is illustrated in Fig. 1 (Top Panel). The preparation of $MoS_2$ flakes is based on chemical exfoliation,[39–41] as described in the methods section; Figures 1 and 2 show examples of the ordered deposition of single flakes on top of metallic nanoholes and 3D gold nanostructures. The method used for the deposition is based on the net mean charge of the $MoS_2$ in solution. As shown in Table1, $MoS_2$ flakes are characterized by a negative surface charge (-39,57 mV ± 0,38 mV) when suspended in deionized water as indicated by the zeta potential (ζ) value measured by Dynamic Light Scattering (DLS). ζ is also known as electrokinetic, which is the potential at the surface-fluid interface of a colloid moving under electric field. Once the charged $MoS_2$ flakes are dispersed in water, ions of the opposite charge will be absorbed at the surface forming a layer of strongly adhered ions (Stern layer) which becomes a diffuse and dynamic layer of a mixture of ions with the increasing distance from the surface. Both Stern and diffuse layer form the electric double layer (EDL) which determines the electrical mobility of the flakes in suspension under electric field and that correspond to the ζ measured by DLS [57]. For this reason, the net surface charge of the flakes reaches almost the neutrality in organic solvent such ethanol (-3,66 ± 0,28 mV) where the EDL is different. This dramatically interferes with the electrophoretic mobility of the $MoS_2$ flakes which is -3,10 ± 0,03 μmcm/Vs and -0,54 ± 0,04 μmcm/Vs measured respectively in water and ethanol (Table1).

**Table1:** Zeta potential and electrophoretic mobility of the $MoS_2$ flakes measured by DLS in water and ethanol. Values are reported as average number (n=3) ± SD.

|  | DI water | Ethanol |
| --- | --- | --- |
| **Zeta potential (mV)** | -39,57 ± 0,38 | -3,66 ± 0,28 |
| **Electrophoretic Mobility (μmcm/Vs)** | -3,10 ± 0,03 | -0,54 ± 0,04 |

The charged nature of the flakes suggests the possible deposition of $MoS_2$ by means of applied voltage, hence realizing an electrophoretic deposition (Fig. 1). The electrophoretic method has been already proposed for $MoS_2$ nanosheet deposition for several applications,[42–44] here, with respect to the previous reported cases, the substrate where the flakes are deposited is not used as electrode. On the contrary, in order to obtain a controlled deposition only over our nanostructures, the substrate is placed in the middle of a fluidic chamber where two Pt electrodes are present. In particular, the nanoholes present in the substrate allow the flow of ions though them and at the same time they represent an obstacle for the $MoS_2$ flakes that are there forced to stop and deposit. The protocol of deposition is the following (illustrated in Fig. 1-Top Panel): 1) the plasmonic holes are prepared on a $Si_3N_4$ membrane (see methods for details on the nanohole fabrication process); 2) the sample to be deposited is first cleaned in oxygen plasma for 60 seconds to facilitate the process; 3) the sample is placed at ground state in a microfluidic chamber equipped with two Pt coated electrodes; 4) $MoS_2$ flakes dispersed in $H_2O$ is injected in the Cis side of the chamber while in Trans side only MilliQ $H_2O$ is present; 5) a suitable voltage is applied for optimized time allowing the electrophoretic deposition; and 6) the chamber is opened and the sample rinsed in EtOH, and the controlled deposition is achieved. In principle this method can be used on every nanostructure comprising a nanohole; in our case, anyway, we observe that the process presents some critical aspects. Firstly we investigated the deposition method on large rectangular arrays of metallic nanoholes. In this configuration, the substrate is a $Si_3N_4$ membrane coated with 100 nm thick gold film where holes of 100 nm diameters were prepared by means of FIB lithography. As can be observed in Fig. 1, the electrophoretic deposition, performed with an applied voltage of 2V for 5 minutes, leads to a well distributed array of $MoS_2$ flakes covering the nanoholes. Anyway, it seems that the holes at the border of the array square did not feel the right ion flow and resulted to be not deposited. This can be due to the electric field configuration inside our chamber and it can be probably overcome increasing the dimension of the system. We also compared this deposition process with a simple drop cast (20 μL of $MoS_2$ in $H_2O$ let dry and rinsed in EtOH) over the same structure. As expected, in this latter case (See SI) the substrate results to be coated randomly even though a large number of holes are still coated. Repeating the electrophoretic deposition process over several identical samples we observe a bad reproducibility in term of quality of the deposited flakes. In fact, also testing several duration of applied voltage, it seems hard to control the number of layers that cover the different nanoholes. As will be reported later, it is possible to obtain over 80 % of metallic holes covered with $MoS_2$ but the number of holes decorated with a single layer flake are low. The evaluation of number of deposited layers will be reported later by means of Raman measurements where the discrimination among single- or few-layer flakes is possible.[45] The evaluation of number of deposited layers will be reported later by means of Raman measurements where the discrimination among single- or few-layer flakes is possible.[45] We tried to investigate the reasons which could lead to the observed low reproducibility. First of all, we compared the ζ values of freshly prepared flakes and flakes stored for several weeks in water and we further measured the surface charge of different $MoS_2$ samples. Results confirmed that the mean net charge is always negative and almost constant around -38 mV (See TableS1 and S2 in SI). Secondly, considering that $MoS_2$ is known to act as catalyst in reduction reactions[58], we investigated if this could potentially interfere with their surface charge. Treating $MoS_2$ with a solution of a strong oxidant ($KMnO_4$) we noticed that, even though the reduction occurs (it is confirmed by the color change of the samples from purple to brownish) the surface charge becomes only slightly more positive, from -38 to -27 mV. This would vary only a little the electrophoretic mobility, therefore the catalytic activity of the $MoS_2$ flakes cannot be associated with the low reproducibility of the deposition (See Table S3 in SI).

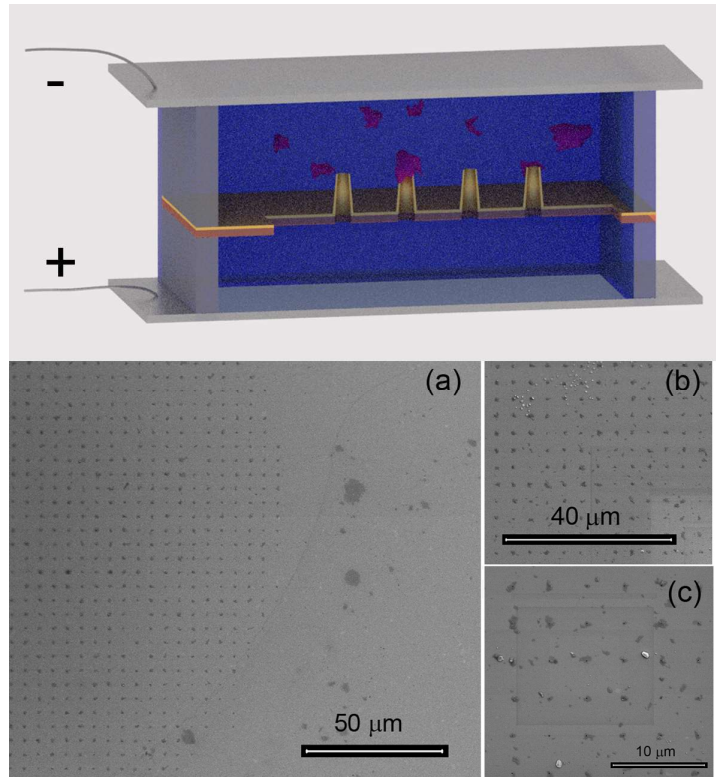

**Figure 1.** Top Panel: Illustration of the electrophoretic process; (a-c) SEM micrographs of gold nanohole arrays decorated with MoS$_2$ flakes by means of electrophoretic deposition.

Once demonstrated the proof-of-concept controlled decoration of nanoholes array, we investigated the applicability of this method to the case of 3D structures (Fig. 2(a-d)). The fabrication of 3D plasmonic structures comprising a nanochannel / nanohole has been demonstrated with different materials and design in several recent papers.[46–48] The main advantage of these particular structures is the possibility to confine and enhance the electromagnetic field (e.m.) in a narrow gap at the apex of the structure. With respect to a nanohole on a flat metallic film the e.m. field enhancement can be significantly higher and the integrated nanochannel can also act as nanofluidic element. In order to illustrate this effect we performed computer simulations (details are reported in the "Methods" section) considering the two structures investigated, i.e. nanoholes and 3D antennas. Fig. 2 illustrates the obtained field confinement / enhancement that can be achieved in the two cases. As can be easily seen in the case of MoS$_2$ flake deposited on top of a 3D antenna (Fig. 2(b)) the electromagnetic field intensity is more confined and can reach higher values respect to the case of a nanohole on a metallic film (Fig. 2(a)).

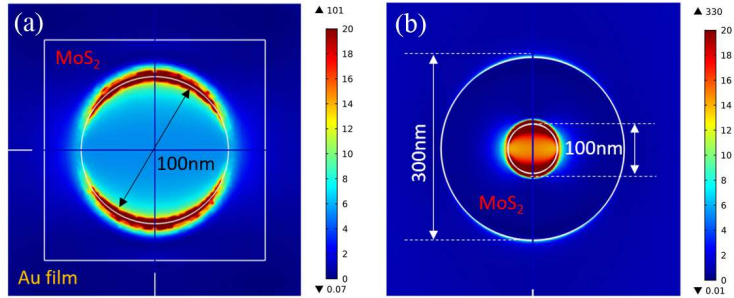

**Figure 2.** Finite Element Method (FEM) simulations of the investigated structures covered with one mono-layer of MoS$_2$. (a) Top view of EM field intensity ($|E/E_0|^2$) in the case of nanohole on metallic (Au) film covered with a MoS$_2$ layer; (b) Top view of $|E/E_0|^2$ in the case of 3D Au antenna covered with a MoS$_2$ layer. In both the case the MoS$_2$ layer covers a 100 nm large hole.

Examples of electrophoretic deposition of Mo$_2$ flakes over these substrates are reported in Fig. 3. As already reported, the vertical antennas used as 3D design can be prepared with different aspect-ratio: from few hundreds of nm high up to few µm. With respect to the case of metallic holes, the electrophoretic force through the higher antennas is not so efficient during the deposition. We expect that higher antennas (See SI) introduce a very high resistivity in the ion flux and, in fact, it was not possible to obtain a reproducible electrophoretic deposition. Decreasing the vertical dimension down to 1 µm and increasing the diameter of the nanochannel (see detail in methods section) it has been possible to obtain the ordered deposition. Anyway, the voltage required is significantly higher respect to the previous case and in order to achieve a high percentage of deposited antennas voltage up to 15 V is needed. Fig. 2(a) and 2(b) illustrate this effect on similar samples. While on the first one an applied voltage of 10 V has been used for 10 minutes, in the second case the voltage has been increased to 15 V leading to a more efficient deposition but with a clear drawback related to the deposition of many flakes also around the 3D antennas. Anyway, in all the cases the deposition results in a partially covering layer that crinkles around the metal in many different ways, ranging from small flakes covering only the top hole to large flakes wrapping the 3D body of the structure.

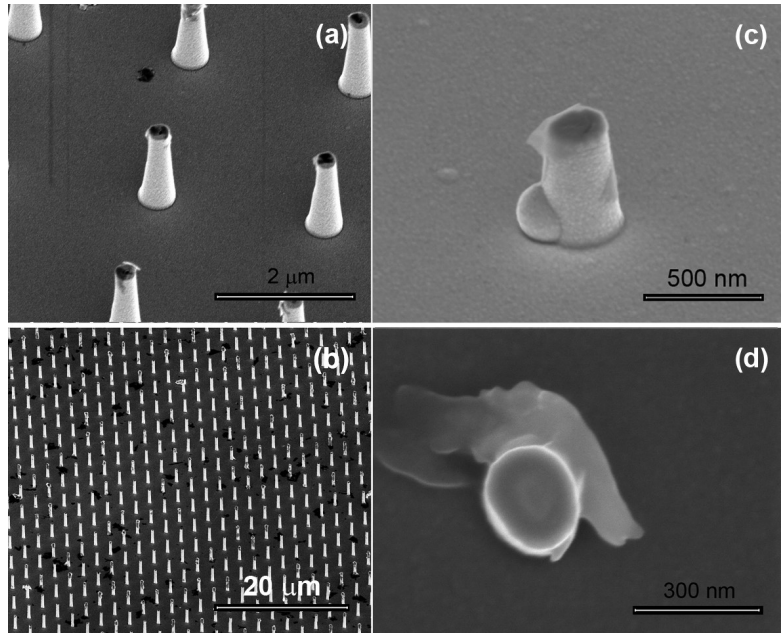

**Figure 3.** SEM micrographs of 3D antennas decorated with $MoS_2$ flakes by means of electrophoretic deposition. (a) 10 V applied voltage, 10 minutes; (b) 15 V applied voltage, 10 minutes; (c) and (d) details of deposited flakes.

To verify the deposition of $MoS_2$ flakes and to evaluate the number of layers deposited over the different positions, we performed Raman spectroscopy on our samples at wavelength of excitation of 532 nm. The measurements were performed by using a Rainshaw InVia Microscope Raman system with a 50 × 0.95 NA objective, collecting the signal with a spectral resolution of 2.5 $cm^{-1}$ and an integration time of 1 second. The system was calibrated by using the intensity of the standard peak at 520 $cm^{-1}$ from a silicon substrate. Figure 4 reports the results of our measurements. Fig. 4(a) illustrates the map over a large arrays of metallic nanoholes while Fig. 4(b) and 4(c) refer to measurements performed on 3D antennas. Raman shifts (excitation wavelength 532 nm) between 400 and 410 $cm^{-1}$ (in correspondence of $A_{1g}$ Raman mode) have been used to evaluate the coverage of the $MoS_2$ over the points. According to the figures a significant difference appears among the samples. In the case of metallic nanoholes, where the deposition results easier also at low applied voltage, the signal appears only in correspondence of the nanostructures. This is a clear demonstration that the deposition strategy covers the desired elements. Over 80 % of the measured holes are decorated with $MoS_2$ flakes. With respect to the case of metallic holes, the maps measured on 3D antennas clearly show what observed during SEM investigations. The sample deposited at 10 V applied voltage is decorated only over half of the antennas, while increasing the voltage to 15 V it is possible to cover almost all the structures but also other part of the substrate results to be deposited.

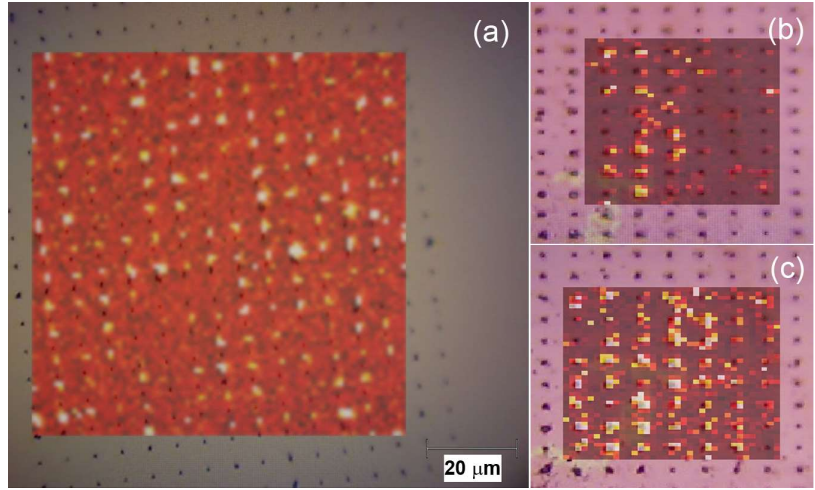

**Figure 4.** Raman maps reporting shifts between 400 and 410 cm$^{-1}$ (in correspondence of A$_{1g}$ Raman mode). (a) Map over 190 nanoholes; (b) map over 3D antennas deposited at 10V for 10 minutes; (c) map over 3D antennas deposited at 15V for 10 minutes.

As previously mentioned, we used Raman spectrum analysis in order to evaluate the number of MoS$_2$ layers deposited on the samples. Figure 5(a)(b) reports the statistical analysis over the different measured points for the metallic nanoholes and 3D nanoantennas samples. As illustrated, the measured points were fitted by Lorentzian functions (Fig. 5(c)). The in-plane (E$^1_{2g}$ at ~ 380 cm$^{-1}$) and out-of-plane (A$_{1g}$ at ~ 404 cm$^{-1}$) Raman modes were always clearly visible and used in the analysis. The difference between the E$^1_{2g}$ and A$_{1g}$ modes (Δf) is known to steadily increase with the number of layers;[7,45,49–51] hence, this parameter can be a reliable quantity to count the number of layers of MoS$_2$. We used Δf to evaluate the percentage of single, double and more layers flakes deposited on the considered array. We consider "1 Layer" Δf is equal to 18 cm$^{-1}$, "2 Layers" when is 22 cm$^{-1}$ and "more layers" when is above 23 cm$^{-1}$. From our analysis, we can conclude that in the case of metallic nanoholes (Fig. 5(a)), a single layer can be hardly deposited (approximately 5 %), while the major part of the flakes are deposited as double (26%) and more than two layers (Fig. 5(b)). This is even more true in the case of 3D nanoantennas where, due to the high voltage needed for the deposition, single layer flakes were deposited in rare case, while the major part of them resulted to be two or more layers. Regardless, we think that this can be improved acting on electrophoretic process conditions and on the exfoliation procedure to obtain better-quality, single-layer flakes in solution. Indeed, herein, we chose to follow a Li-intercalation protocol that with respect to liquid-phase exfoliation, can provide stable suspensions without any surfactant (that may hamper both anchoring and plasmonic behavior) and with higher exfoliation degree,[52] which, in our case, was maximized by doubling the Li dose. Unfortunately, MoS$_2$ strongly tends to break up during the exfoliation, resulting in a quite large size dispersion. Even if our deposition procedure was proved to work with all nanosheet sizes, we believe that improving the synthetic procedure to have flakes with homogenous thickness and a controlled lateral dimension would allow further optimization of the deposition parameters and, consequently, enhancement of the performances of these hybrid systems. Finally, although our synthetic procedure is highly time consuming and low yielding, Li-exfoliation may be scaled-up by switching from chemical to electrochemical intercalation.[53,54] This process would allow the preparation of single-layer $MoS_2$ with higher throughput, which is necessary for the application of these types of systems on a large scale.

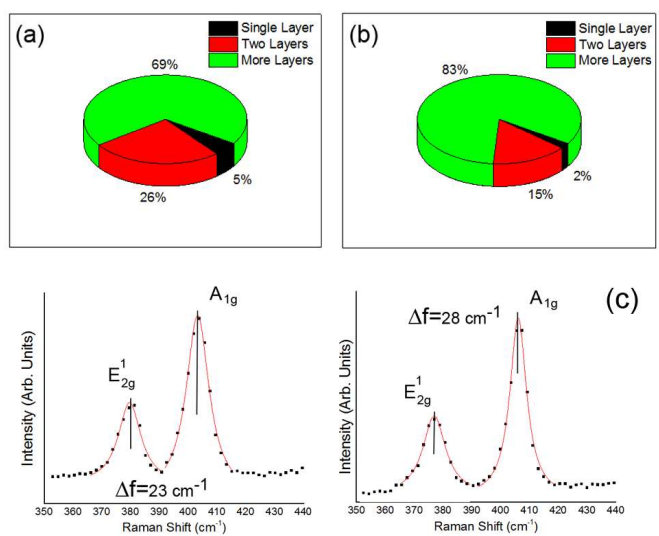

**Figure 5**. Statistical analyses based on Raman spectrum to evaluate the number of deposited $MoS_2$ layers. (a) Metallic nanoholes deposited at 2V; (b) 3D nanoantennas deposited at 15V; (c) examples of analyzed spectrum.

**Conclusions**

In summary, we presented a method for hybrid plasmonic 2D material structure preparation. The fabrication procedure allows to obtain ordered structures over large array using a low-cost procedure and without the use of complex lithographic processes. This strategy can be applied to not only $MoS_2$ but also many 2D materials for which their net charge of their surface can be used to guide the electrophoretic process.

We believe that such an approach can be interesting for the realization of new hybrid devices for use in several applications, including photoluminescence, strong coupling and valley-polarization studies. With respect to previously reported hybrid plasmonic nanostructures, our scheme significantly reduces the complexity of fabrication, leading to a more robust and low-cost approach for the integration of 2D materials with plasmonic nanopores.

**Methods**

Exfoliation of $MoS_2$

In a glove-box (water < 1 ppm, $O_2$ < 10 ppm), $LiBH_4$ (0.109 g, 5 mmol) and $MoS_2$ (0.320 g, 2 mmol) were grounded in a mortar and subsequently transferred in a Schlenk-tube, which then was brought outside of the glove-box and connected to a Schlenk-line. This mixture was heated in a sand bath at 330°C for 4 days under nitrogen. Afterwards, the Schlenk-tube was brought again inside the dry-box, where it was newly grounded with additional $LiBH_4$ (0.109 g, 5 mmol). The sample was subsequently heated for 3 days at 330°C under nitrogen.

The intercalation product was added in a single shot in 270 ml of degassed water and the resulting suspension was bath-sonicated for 1 h to facilitate the exfoliation.

In order to remove the LiOH produced, the suspension was equally divided into six centrifugation tubes (45 ml/tube) and centrifuged at 10000 rpm ($23478g$) for 20 min for three times, replacing the supernatant with clean solvent.

To select the flake size, the suspension purified by LiOH, was progressively centrifuged at 8000 ($15026g$ – 8K fraction), 6000 ($8452g$ – 6K), 4000 ($3757g$ – 4K), 3000 ($2113g$ – 3K), 2000 ($939g$ – 2K) and 1000 rpm ($235g$ – 1K), by collecting each time the top 2/3 of the supernatant and by replacing it with ultrapure water.

Fabrication of plasmonic nanostructures

The fabrication of the metallic nanoholes follows simple and robust procedures for the 2D and 3D geometry. In both cases the substrate was a $Si_3N_4$ membrane (100 nm thick) prepared on a Silicon chip. The 2D holes have been prepared by means of FIB milling with a current of 80 pA. After the milling a thin layer of gold, ca. 100 nm, has been deposited on the top side of the membrane. The fabrication of 3D nanoholes array follows the procedure illustrated in Ref. 24. A thin layer of S1813 optical resist is spinned on top of the membrane with a final thickness equal to the height of the structure we want to obtain. This layer is exposed by secondary electron during the FIB milling of the nanoholes from the bottom of the membrane, and after the development in acetone and a final metal deposition (ca. 40 nm) a 3D hollow antenna is obtained. For these experiment 3D antennas with inner diameter between 30 and 100 nm and final heigth between 1000 and 2000 nm have been prepared.

FEM Simulation of MoS$_2$ flakes integrated over plasmonic nanostructures

We investigate the plasmonic properties of the structure by means of finite element method (FEM) simulations by using of the RF Module in Comsol Multiphysics taking into account the geometries that can actually be fabricated. Here we consider a hole of 100 nm diameter into Si3N4//Au 100//50 nm membrane. In the case of 3D antennas we consider an antenna with height equal tio 1000n m and, inner and outer diameter of 100 nm and 300 nm, respectively. A top single layer of MoS2 has been considered in both the cases.

The refractive indexes of Au and MoS$_2$ were taken from the works of Rakić *et al* and Zhang *et al*.[55,56] The model computes the electromagnetic field distribution in each point of the simulation region, enabling the extraction of the quantities plotted in Figure 2. The unit cell was set to be 400 nm wide in both x- and y-directions, with perfect matching layers (150 nm thick) at the borders. A linearly polarized incident plane wave was assumed to imping on the structure from the air side.

**Acknowledgements**
The research leading to these results has received funding from the Horizon 2020 Program, FET-Open: PROSEQO, Grant Agreement n. [687089].

**Author Contributions**
D.M. worked on the MoS$_2$ preparation and characterization; A.J. performed the electrophoretic deposition; P.P. fabricated the structures; G.G. performed DLS characterizations; N.M. performed the simulations; M.D. helped during the manuscript preparation; P.V. S.A., F.D.A. and D.G. coordinated the work.

**Competing financial interests:**
The authors declare no competing financial and non-financial interests.